\documentclass[final,5p,times]{elsarticle}

\biboptions{sort&compress,numbers}

\usepackage{amsmath}

\usepackage{graphicx}

\usepackage{textcomp} 
\usepackage{color}

\usepackage[
colorlinks,
citecolor=blue,
]{hyperref}

\begin{document}

\begin{frontmatter}

\title{First measurement of a long-lived $\pi^+ \pi^-$ atom lifetime\\[3ex]
{\normalsize \it 
This paper is dedicated to the memory of our colleague Valeriy Yazkov \\[-1ex]
who could not anymore see the full impact and significance of 
his contributions to the DIRAC experiment.}
}

\author[s]{B.Adeva} 
\author[d]{L.Afanasyev} 
\author[itm]{A.Anania} 
\author[b]{S.Aogaki} 
\author[zu]{A.Benelli} 
\author[p]{V.Brekhovskikh} 
\author[cz]{T.Cechak} 
\author[jt]{M.Chiba} 
\author[p]{P.V.Chliapnikov} 
\author[cz]{P.Doskarova}
\author[c]{D.Drijard} 
\author[d]{A.Dudarev}
\author[b]{D.Dumitriu}
\author[b]{D.Fluerasu}
\author[p]{A.Gorin} 
\author[d]{O.Gorchakov} 
\author[d]{K.Gritsay} 
\author[if]{C.Guaraldo} 
\author[b]{M.Gugiu} 
\author[c]{M.Hansroul} 
\author[czr]{Z.Hons} 
\author[zu]{S.Horikawa} 
\author[jku]{Y.Iwashita} 
\author[d]{V.Karpukhin} 
\author[cz]{J.Kluson} 
\author[k]{M.Kobayashi} 
\author[d]{V.Kruglov} 
\author[d]{L.Kruglova} 
\author[d]{A.Kulikov} 
\author[d]{E.Kulish} 
\author[itm]{A.Lamberto} 
\author[c,u]{A.Lanaro} 
\author[cza]{R.Lednicky} 
\author[s]{C.Mari\~nas}
\author[cz]{J.Martincik} 
\author[d,c]{L.Nemenov} 
\author[d]{M.Nikitin} 
\author[jks]{K.Okada} 
\author[d]{V.Olchevskii} 
\author[v]{V.Ovsiannikov} 
\author[b]{M.Pentia} 
\author[it]{A.Penzo} 
\author[s]{M.Plo} 
\author[cz]{P.Prusa} 
\author[itm]{G.F.Rappazzo} 
\author[s]{A.Romero Vidal} 
\author[p]{A.Ryazantsev} 
\author[p]{V.Rykalin} 
\author[s]{J.Saborido} 
\author[be]{J.Schacher\corref{cor1}} 
\ead{Juerg.Schacher@cern.ch} 
\author[p]{A.Sidorov} 
\author[cz]{J.Smolik} 
\author[jks]{F.Takeutchi} 
\author[cz]{T.Trojek} 
\author[m]{S.Trusov} 
\author[cz]{T.Vrba}
\author[m]{V.Yazkov\fnref{yaz}} 
\author[k]{Y.Yoshimura} 
\author[d]{P.Zrelov} 

\cortext[cor1]{Corresponding author} 
\fntext[yaz]{deceased}

\address[s]{Santiago de Compostela University, Spain}
\address[d]{JINR Dubna, Russia}
\address[itm]{INFN, Sezione di Trieste and Messina University, Messina, Italy}
\address[b]{IFIN-HH, National Institute for Physics and Nuclear Engineering, 
Bucharest, Romania}
\address[cz]{Czech Technical University in Prague, Prague, Czech Republic}
\address[p]{IHEP Protvino, Russia}
\address[jt]{Tokyo Metropolitan University, Japan}
\address[c]{CERN, Geneva, Switzerland}
\address[if]{INFN, Laboratori Nazionali di Frascati, Frascati, Italy}
\address[czr]{Nuclear Physics Institute ASCR, Rez, Czech Republic}
\address[zu]{Zurich University, Switzerland}
\address[jku]{Kyoto University, Kyoto, Japan}
\address[k]{KEK, Tsukuba, Japan}
\address[u]{University of Wisconsin, Madison, USA} 
\address[cza]{Institute of Physics ASCR, Prague, Czech Republic}
\address[jks]{Kyoto Sangyo University, Kyoto, Japan}
\address[v]{Voronezh State University, Russia}
\address[it]{INFN, Sezione di Trieste, Trieste, Italy} 
\address[be]{Albert Einstein Center for Fundamental Physics, 
Laboratory of High Energy Physics, Bern, Switzerland}
\address[m]{Skobeltsin Institute for Nuclear Physics of Moscow State University, 
Moscow, Russia}



\begin{abstract}

The adapted DIRAC experiment at the CERN PS accelerator observed for 
the first time long-lived hydrogen-like $\pi^+\pi^-$ atoms, 
produced by protons hitting a beryllium target. A part of these atoms 
crossed the gap of 96~mm and got  broken up in 
the 2.1~\textmu{}m thick platinum foil. 
Analysing the observed number of atomic pairs, 
$n_A^L= \left.436^{+157}_{-61}\right|_\mathrm{tot}$, 
the lifetime of the 2$p$ state is found to be 
${\tau_{2p}=(\left.0.45^{+1.08}_{-0.30}\right|_\mathrm{tot}) \cdot10^{-11}}$s, 
not contradicting the corresponding QED $2p$ state lifetime 
${\tau_{2p}^\mathrm{QED}=1.17 \cdot 10^{-11}}$s. 
This lifetime value is three orders of magnitude larger than 
our previously measured value of the $\pi^+\pi^-$ atom ground state lifetime 
$\tau=(\left.3.15^{+0.28}_{-0.26}\right|_\mathrm{tot})\cdot 10^{-15}$s. 
Further studies of long-lived $\pi^+\pi^-$ atoms will allow 
to measure energy differences between $p$ and 
$s$ atomic states and so to determine $\pi\pi$ scattering lengths with 
the aim to check QCD predictions. 

\end{abstract}

\begin{keyword}

DIRAC experiment
\sep
double-exotic atom
\sep
pionium
\sep
lifetime
\sep
QCD

\end{keyword}

\journal{Physics Letters B}

\end{frontmatter}

\section{Introduction}
\label{sec:intro}
 
The DIRAC collaboration aims to check low-energy QCD predictions 
using double-exotic $\pi^+\pi^-$ and $\pi^{\pm} K^{\mp}$ atoms, 
which have been observed and studied 
\cite{AFAN93,ADEV05,ADEV11,ADEV09,ADEV14,ADEV16,ADEV17}. 
In strong inclusive processes, these atoms are produced in 
$s$ states distributed according to $n^{-3}$, $n$ being 
the principal quantum number \cite{NEME85}.

The decay probability of short-lived $\pi^+\pi^-$ atoms 
($A_{2\pi}$, pionium) in $s$ states is dominated (99.6\%) by 
the annihilation process \cite{URET61,BILE69,JALL98,IVAN98,GASS01} 
$\pi^{+} + \pi^{-} \rightarrow \pi^{0} + \pi^{0}$
and is given by the $\pi\pi$ $s$-wave scattering lengths combination 
$|a_0-a_2|$ ($a_I$ is the $\pi\pi$ scattering length for isospin $I$):

\begin{equation}\label{eq:tau-aa}
\frac{1}{\tau} \approx \Gamma(A_{2\pi} \rightarrow \pi^{0} \pi^{0}) = 
R \left| a_0-a_2 \right|^2 \quad \mbox{with} \quad 
R \propto \left|\psi_{nl}(0)\right|^2.
\end{equation}
The expression $\psi_{nl}(0)$ is the pure Coulomb atomic wave 
function at the origin with principal quantum number $n$ 
and orbital quantum number $l$. The precision of the ratio $R$ 
is 1.2\% \cite{GASS01}, and the scattering lengths 
and their combination have been calculated in 
low-energy QCD \cite{COLA01}.
The DIRAC experiment has measured these lifetimes and so derived 
combined $\pi\pi$ \cite{AFAN93,ADEV05,ADEV11} and for the first time 
$\pi K$ \cite{ADEV14,ADEV17} scattering lengths.

After investigation of $\pi \pi$ and $\pi K$ atoms with 
``short'' lifetimes of order $10^{-15}$s, 
DIRAC continues to study $\pi \pi$ 
atomic states ($A_{2\pi}^L$) of ``long'' lifetimes of 
order $10^{-11}$s as considered in Ref. \cite{NEME85}. 
Moving inside the target, relativistic short-lived $A_{2\pi}$ 
interact with the electric field of the target atoms, 
resulting in a possible change of 
the $A_{2\pi}$ orbital momentum $l$ by one or more units. 
Some ($N_A^L$) of them ($N_A$) will leave the target with $l>0$. 
For such states, $\left|\psi_{nl}(0)\right|^2=0$ and thus all decays are 
suppressed in accordance with (\ref{eq:tau-aa}).
Therefore, the decay mechanism 
of such excited states is the radiative de-excitation to 
an $ns$ state, annihilating subsequently with the ``short'' lifetime 
$\tau \cdot n^3$ into two $\pi^0$.  
The shortest $A_{2\pi}^L$ lifetime is the 2$p$ state lifetime 
$\tau_{2p} \approx \tau_{2p}^\mathrm{QED}=1.17 \cdot 10^{-11}$~s, 
which is three orders of magnitude larger than 
the minimum $A_{2\pi}$ lifetime. 
For an average momentum of detected 
$A_{2\pi}$ $\langle p_A \rangle$=4.44~GeV/$c$ ($\gamma \simeq 15.9$), 
the decay lengths are 5.6~cm (2$p$), 18.4~cm (3$p$), 43~cm (4$p$), 
83~cm (5$p$) and 143~cm (6$p$).

These large atom decay lengths in the laboratory (lab) system 
open a possibility to measure the energy splitting between $ns$ and $np$ 
levels, which depends on an other combination of $\pi\pi$ scattering lengths,  
$2a_0+a_2$ \cite{NEME01,NEME02}. 

In this paper, the DIRAC collaboration presents 
the first lifetime measurement of long-lived $\pi^+\pi^-$ atoms, 
observed by DIRAC as described in \cite{ADEV15}.

\section{Setup for investigation of long-lived $\pi^+\pi^-$ atoms}
\label{sec:setup}

\begin{figure}[tb]
  \begin{center}
  \includegraphics[width=0.92\columnwidth]{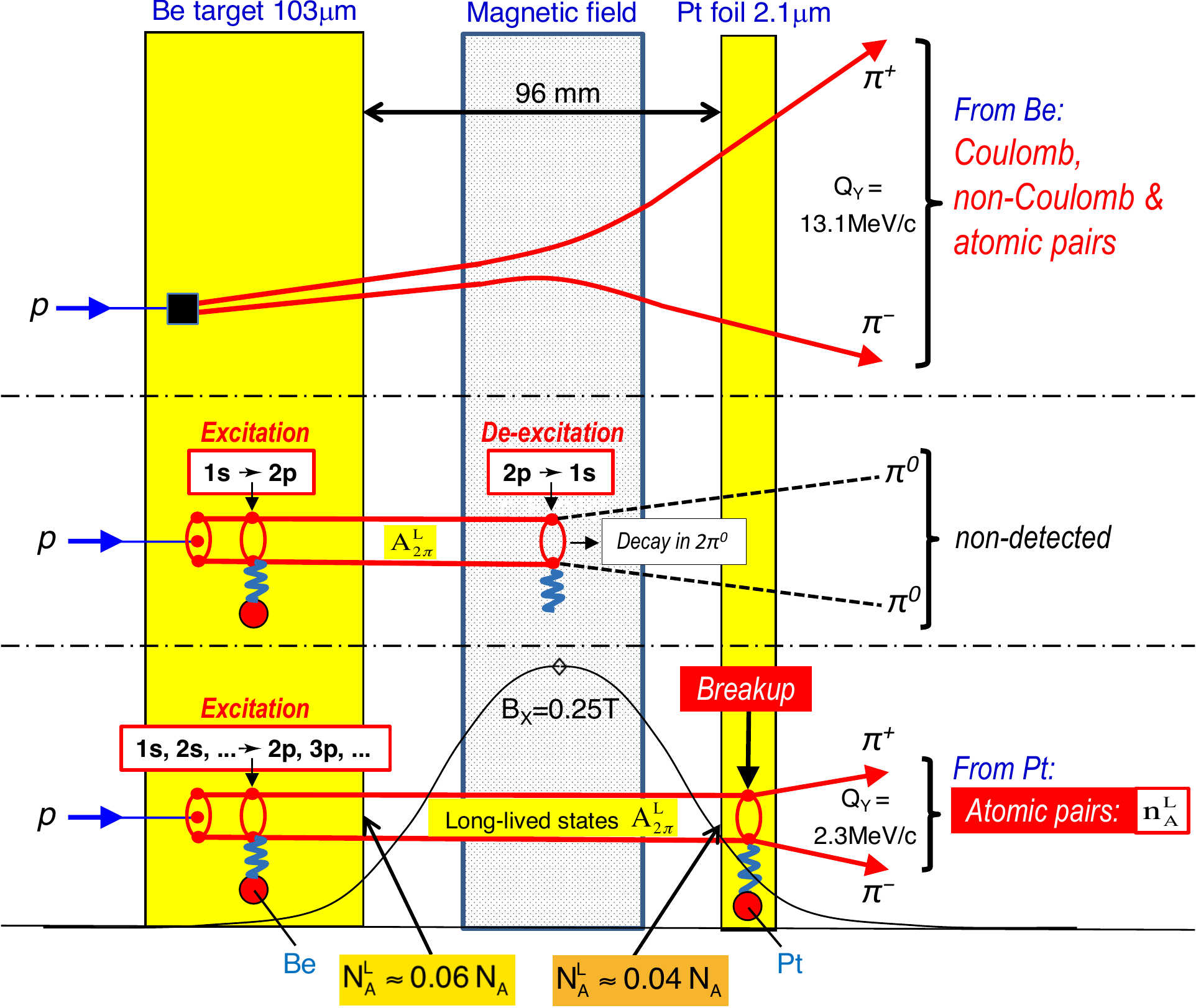}
  \caption{
  Method to investigate long-lived $A_{2\pi}^L$ by 
  means of a breakup foil (side view). $N_A$ is the number 
  of the totally produced $A_{2\pi}$ and $N_A^L$ 
  the number of excited $A_{2\pi}^L$ with 
  $l>0$.}
  \vspace{-2ex}
  \label{fig:1}
  \end{center}
\end{figure}

\begin{figure}[tb]
  \begin{center}
  \includegraphics[width=0.99\columnwidth]{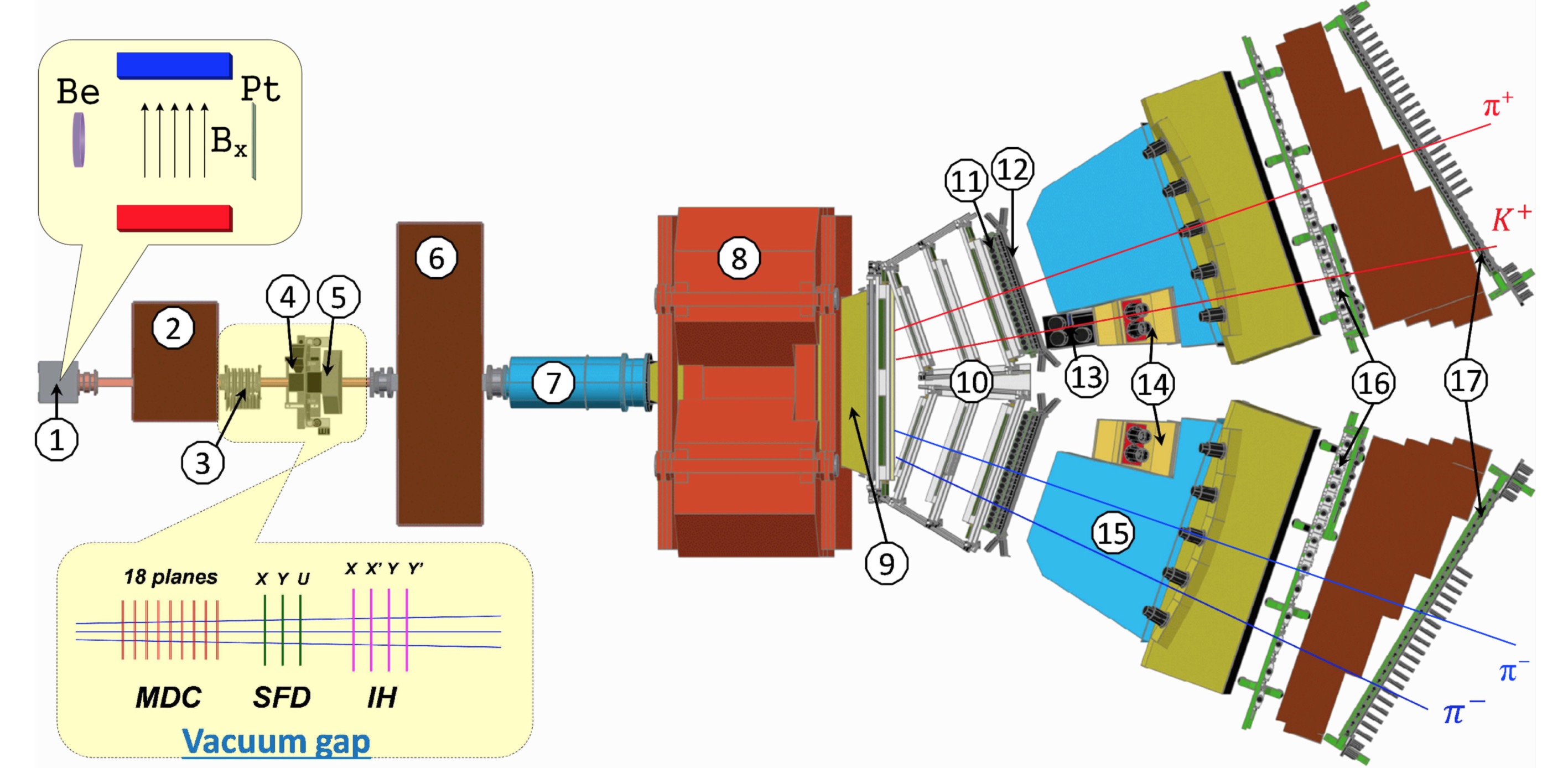}
  \caption{
  Top view of the DIRAC setup:  
  1 -- target station with insertion, showing the Be target, 
  magnetic field and Pt breakup foil;
  2 -- first shielding;
  3 -- microdrift chambers (MDC);
  4 -- scintillating fiber detector (SFD); 
  5 -- ionisation hodoscope (IH); 
  6 -- second shielding; 
  7 -- vacuum tube; 
  8 -- spectrometer magnet; 
  9 -- vacuum chamber; 
  10 -- drift chambers (DC); 
  11 -- vertical hodoscope (VH); 
  12 -- horizontal hodoscope (HH); 
  13 -- aerogel Cherenkov; 
  14 -- heavy gas Cherenkov; 
  15 -- nitrogen Cherenkov; 
  16 -- preshower (PSh); 
  17 -- muon detector. 
  (The plotted symmetric and asymmetric events are 
  a $\pi \pi$ and $\pi K$ pair, respectively.)
  }
  \label{fig:2}
  \end{center}
  \vspace{-3ex}
\end{figure}

To investigate long-lived $A_{2\pi}^L$ atoms, the primary 24~GeV/c CERN PS 
proton beam with intensity of 3.0 $\cdot 10^{11}$ protons per spill 
\cite{SPSC11}  hits a 103~\textmu{}m thick Be target 
(Figures~\ref{fig:1} and \ref{fig:2}). 
On their way to
the 2.1~\textmu{}m thick Pt foil at a distance of 96~mm behind the Be target, 
a part of the produced long-lived atoms $A_{2\pi}^L$, 
depending on their lifetimes, decays, whereas 
the other part enters the Pt foil. 
By interacting with Pt atoms, the atoms $A_{2\pi}^L$ break up 
(get ionized) and generate atomic $\pi^+\pi^-$ pairs with 
relative momenta $Q<3$~MeV/$c$ in their pair 
centre-of-mass (c.m.) system. 
The foil is introduced at 7.5~mm above 
the primary proton beam to avoid interaction of 
the beam halo with Pt. The beam position in 
the vertical plane was permanently monitored. 
Between target and 
breakup foil, there was installed a permanent retractable magnet 
\cite{note1304} with a pole distance of 60~mm (insertion of 
Figure~\ref{fig:2}) and a maximum horizontal field strength of 0.25~T 
(bending power 0.02~Tm). The magnet enlarges the value of 
the vertical component $Q_{Y}$ of $\pi^+\pi^-$ pairs, generated 
in Be, on $Q_Y=13.15$~MeV/$c$ (see Figure~\ref{fig:1}). For simulated events, 
this value is $Qy=13.12$~MeV/$c$.
For all pairs generated in Pt foil, the fringing magnetic field modifies the 
vertical component $Q_{Y}$ only by 2.3~MeV/$c$.
This large difference in the $Q_Y$ shift for pairs from Be and Pt 
allows to suppress the background by a factor of about 6 in 
detecting atomic pairs from long-lived $A_{2\pi}^L$. 
For $\mathrm{e}^+\mathrm{e}^-$ pairs generated in Be, the peak is shifted 
to $Q_Y=12.9$~MeV/$c$ and its position used to measure the magnetic field 
stability during the 6 month data taking \cite{ADEV15}.

The DIRAC setup \cite{DIRA16}, sketched in Figure~\ref{fig:2}, 
identifies pions, kaons, protons, electrons and muons by means of 
Cherenkov, preshower and muon detectors and time-of-flight measurement. 
The achieved resolution in the particle momentum is 
$\Delta p/p \simeq 3\cdot 10^{-3}$ and the precision of the $Q$ components 
$\sigma_{QX} \approx \sigma_{QY} \approx 0.44~\rm{MeV}/c$ and  
$\sigma_{QL} \approx 0.50~{\rm MeV}/c$. 
This high resolution enables the extraction of 
an atom signal in form of small $Q$ atomic pairs. 
The secondary channel with the solid angle 
$\Omega = 1.2 \cdot 10^{-3}$~sr is vertically inclined relative 
to the proton beam by $5.7^\circ$ upward. 

The events are reconstructed by means of the 
DIRAC $\pi\pi$ analysis software described in \cite{ADEV15}, 
exploiting information from DC, SFD, IH and the proton beam position 
on the target. The setup is aligned using properties of $\Lambda$ 
($\bar{\Lambda}$) decays \cite{note1303}.

\section{Generation of short-lived $\pi^+\pi^-$ atoms,
Coulomb and non-Coulomb pairs}
\label{sec:gen-short}

Coulomb $\pi^+\pi^-$ pairs are produced either directly or 
originating from short-lived (e.g. $\Delta$, $\rho$) 
or medium-lived (e.g. $\omega$, $\phi$).   
These pairs undergo Coulomb final state interaction (FSI) 
resulting in modified unbound states or atoms. 
Pairs from long-lived sources (e.g. $\eta'$, $\eta$), 
called non-Coulomb, are nearly unaffected by 
Coulomb interaction. The accidental pairs arise from 
different interactions. 

The cross section of $\pi^+\pi^-$ atom production is expressed via 
the inclusive production cross section of $\pi^+\pi^-$ pairs 
from short-lived sources without FSI, multiplied by 
$\left|\psi_{nl}(0)\right|^2$ \cite{NEME85}. 
The latter factor differs from zero only for $s$ states. Thus, 
atoms are only produced in $s$ states with probabilities dictated by 
the $n$-dependence of $\left|\psi_{nl}(0)\right|^2$ as $n^{-3}$: 
$W_1=83.2\%$, $W_2=10.4\%$, $W_3=3.1\%$, $W_{n>3}=3.3\%$.

In complete analogy, the production of $\pi^+ \pi^-$  Coulomb pairs 
is expressed via the inclusive production cross section of 
$\pi^+\pi^-$ pairs without FSI, modified by 
the Coulomb enhancement function $A_C(q)$, which depends on 
the pair relative momentum $q$ ($\sim1/q$ for small $q$). 
This function, the well-known Gamov-Sommerfeld-Sakharov factor 
\cite{GAMO28, SOMM31, SAKH91}, is used for simulation of Coulomb pairs. 

The relative yield between atoms and Coulomb pairs \cite{AFAN99} 
is expressed by the ratio of the above cross sections. 
Thus, the total number $N_A$ of produced $A_{2\pi}$ is 
determined via the number of Coulomb pairs $N_C$ by 
the model-independent relation: 
\begin{equation}\label{eq:number_A}
N_A = K(q_0) N_C(q \le q_0) 
\quad \mathrm{with} \quad
K(2~\mathrm{MeV}/c) = 0.615 \,,
\end{equation}
where $N_C(q \le q_0)$ is the number of Coulomb pairs with $q \le q_0$  
and $K(q_0)$ a known function of $q_0$ \cite{AFAN99}. 

In the above consideration, the production of $\pi^+\pi^-$ pairs 
and their FSI are assumed to be pointlike. Contributions from 
medium-lived resonances \cite{LEDN08,note1205} are taken into account as 
small corrections to the production cross sections.

\section{Production of long-lived $\pi^+\pi^-$ atoms}
\label{sec:long-lived}

The short-lived $\pi^+\pi^-$ atoms, which propagate in 
the Be target after their production in $ns$ states, are 
either decaying or interacting with Be. This interaction with 
the electric field of target atoms will exite/de-excite 
or break up $\pi^+\pi^-$ atoms. In the excitation/de-excitation 
processes, only transitions with $Z$-parity ($P^Z=(-1)^{l+m}$)  
conservation are permitted \cite{TARA91}, changing 
the orbital momentum by one unit or more and so forming 
long-lived atoms. The quantization axis is 
along the lab atom momentum. 
The lifetime of $\pi^+\pi^-$ atoms in the lab system depends 
on the quantum numbers $n$,$l$ and the atom momentum. 

\begin{figure}[tb]
\begin{center}
\includegraphics[width=0.9\columnwidth]{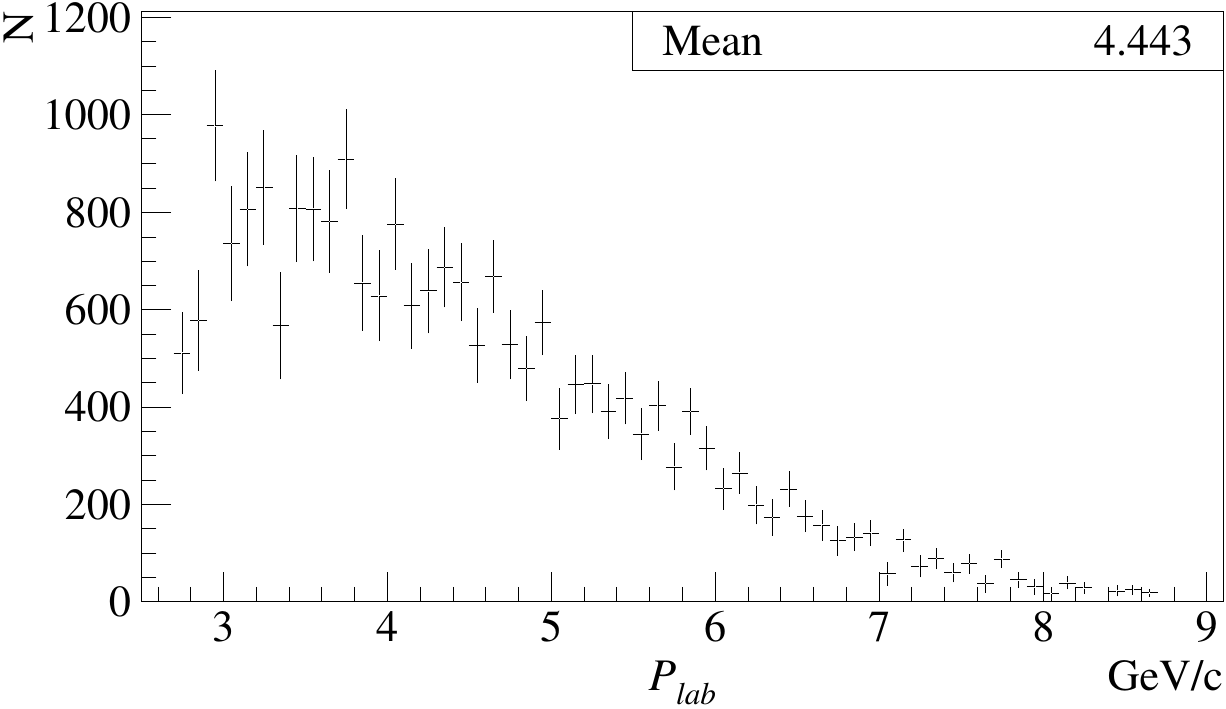}
\caption{Lab momentum spectrum of $A_{2\pi}$ produced in Be, 
as used in the simulation.} 
\label{fig:3}
\end{center}
\end{figure}

The populations of all atomic states with quantum numbers $n$, $l$, $m$,  
as a multi-level quantum system, are described by an infinite set of  
transport equations in terms of probabilities \cite{AFAN96,AFAN04}. 
These equations account for excitation/de-excitation and annihilation   
processes and describe the atomic state populations from 
the production point up to the target exit. For the calculations below, 
our measured value of the $A_{2\pi}$ ground state lifetime  
$\tau=3.15\cdot 10^{-15}$s \cite{ADEV11} and the spectrum of 
the atom lab momentum as extracted from the experimental distribution 
of the Coulomb pairs (Figure~\ref{fig:3}) are used. 

\begin{figure}[tb]
\begin{center}
\includegraphics[width=0.9\columnwidth]{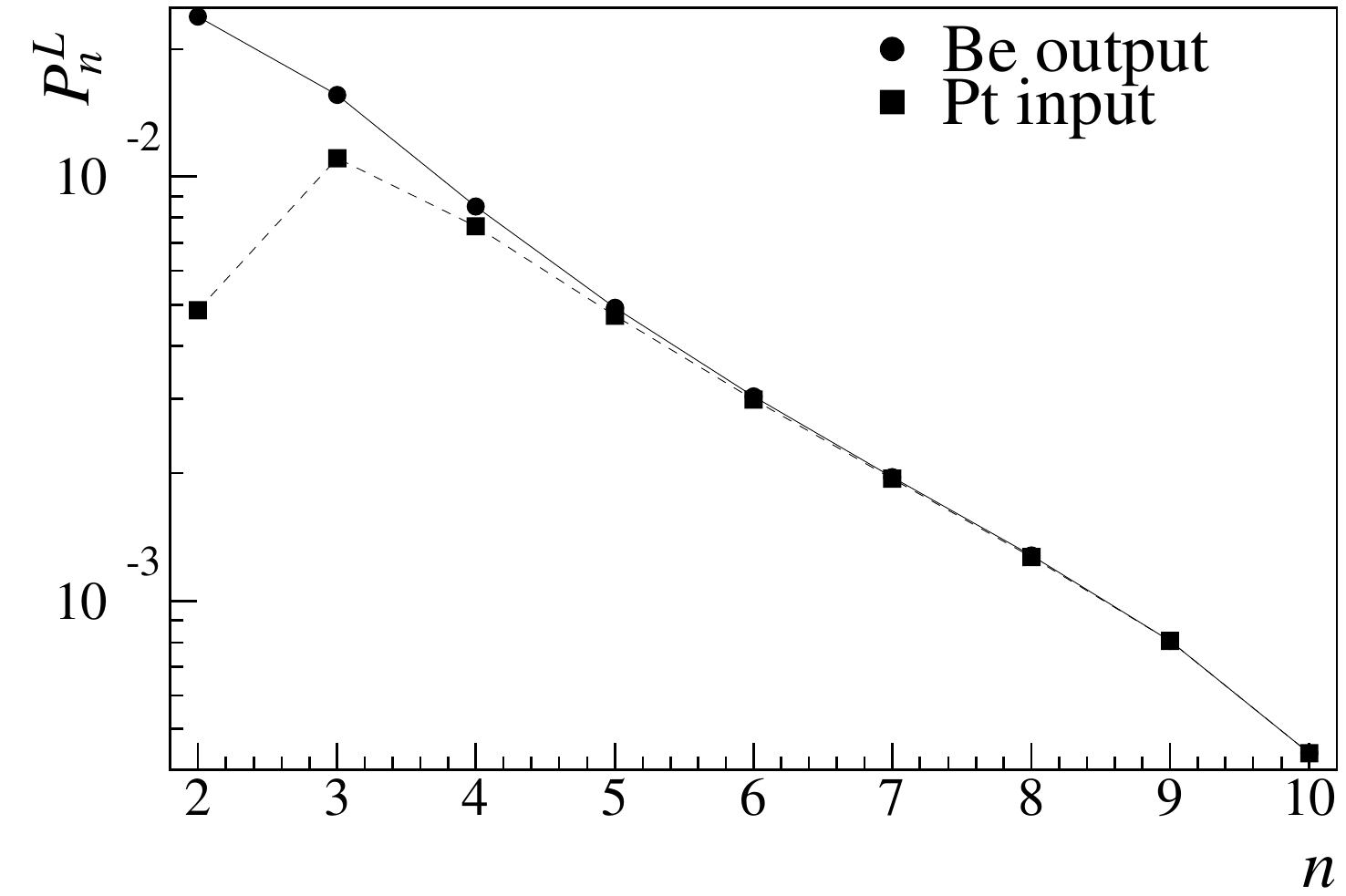}
\caption{Populations $P_n^L$ of long-lived states $A_{2\pi}^L$ versus $n$, 
summed over $l$ and $m$, at the exit of the Be target ($\bullet$) 
and at the Pt foil entry ($\blacksquare$).} 
\label{fig:4}
\end{center}
\vspace{-3ex}
\end{figure}

\subsection{Population of long-lived states}
\label{sec:ll-pop}

In this paper, the transport equations are solved numerically 
for all states with principal quantum numbers $n\leq 10$. The populations of 
states with higher $n$ are taken into account by an extrapolation procedure.  
Only total and excitation/de-excitation cross sections are considered: 
the cross sections are calculated in the first Born approximation 
(one-photon exchange), using Moli\`{e}re parametrization of 
the Thomas-Fermi formfactors for Be and Pt atoms. 

Figure \ref{fig:4} shows the atomic state population $P_n^L$ versus $n$,
summed over $l$ and $m$, at the exit of the Be target and before the Pt foil. 
The almost pure exponential behaviour at high $n$ allows 
to extrapolate to $n>10$ and to evaluate the population of 
the non-negligible infinite ``tail''. 
Thus, the population of all discrete states can be estimated. 
The population at $n=10$ is excluded from the extrapolation procedure, 
as this population is underestimated because of cutting 
the infinite set of transport equations, excluding 
de-excitations to this state from higher $n$ states. 
The cut does almost not affect the populations with smaller $n$, 
as transitions between nearest $n$ levels dominate. 
A fitting procedure over all points is badly suited for 
the ``tail'' estimation, as the $n$ dependence is mainly dictated 
by two different dependences: 
production rates as $n^{-3}$ and total cross sections roughly as $n^2$. 
The following two $n$ dependences are chosen for extrapolation: 
exponential, $a\exp{(-bn)}$, and hyperbolic, $cn^{-3}+dn^{-5}$. 
The free parameters $a$, $b$ and  $c$, $d$ have been calculated 
for $n$, using the populations $P_n^L$ and $P_{n-1}^L$.  

\begin{figure}[tbp]
\begin{center}
\includegraphics[width=0.9\columnwidth]{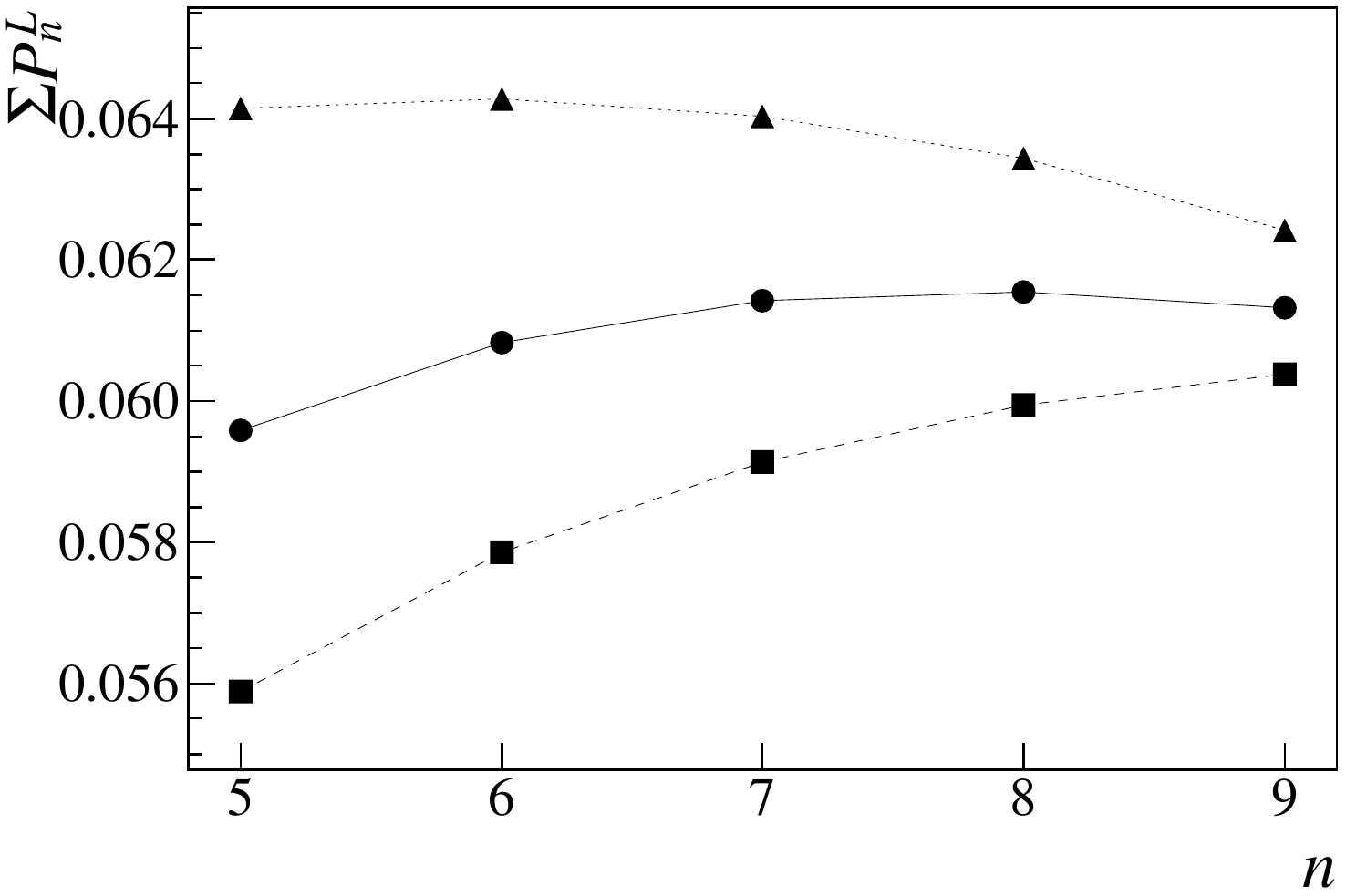}
\caption{Summed populations $\sum P_n^L$ of all long-lived atomic states at 
the exit of the Be target as a function of $n$ used for ``tail'' estimation. 
For each $n$, the two upper curves show the sum of state 
populations for the given $n$ plus different ``tail'' estimations 
calculated from populations for $n$ and $n-1$ 
(exponential ``tail'' -- $\bullet$, hyperbolic ``tail'' -- $\blacktriangle$). 
The lower curve ($\blacksquare$) presents the sum of the population for 
the given $n$ plus the population for $n+1$ instead of the ``tail''. }
\label{fig:5}
\end{center}
\vspace{-3ex}
\end{figure}

Figure \ref{fig:5} illustrates the ``tail'' estimations.
The numerical solution of the transport equations are obtained for 
$n\leq 10$. Then, the populations at $n=9$ and 8 are used for 
the ``tail'' estimation. The two upper points at $n=9$ in 
Figure~\ref{fig:5} show the sum of the populations for $n\leq 9$ 
plus the different ``tail'' estimations 
(exponential -- $\bullet$, hyperbolic -- $\blacktriangle$). 
The lower point ($\blacksquare$) presents the sum of the populations 
with $n\leq 9$ plus the population for $n=10$ instead of the ``tail''. 
The curves are obtained by applying the same procedure to the points 
with lower $n$. The exponential ``tail'' estimation is chosen 
as the mean value for the population of long-lived states, 
as it has the smallest slope and is nearly flat for $n\geq7$. 
The two other curves are used as error band. 
The illustrative numerical values are calculated for 
the average atom momentum $\langle p_A \rangle=4.44~$GeV/$c$ 
(Figure~\ref{fig:3}): 
the sum of the long-lived state population at the Be exit is 6.04\% 
for $n\leq 10$ (5.99\% for $n\leq 9$) of the total number $N_A$ 
of produced $A_{2\pi}$; 
the exponential ``tail'' $n\geq10$ is 0.14\%,  
the hyperbolic ``tail'' 0.25\%, and 
the population of all long-lived states at the Be exit is given as 
$N_A^{L,\mathrm{Be}}=(6.13^{+0.11}_{-0.09}) \cdot 10^{-2}\times N_A$. 
Moreover, about 17\% of $A_{2\pi}$ are produced in short-lived states ($ns$) 
and annihilate just behind the Be target. About 4.6\% of the atoms break up 
in Be producing atomic pairs. Let us underline that accuracy of all 
above numbers accounts for the accuracy of the extrapolation procedure. 

As mentioned above, the cut of the infinite set of transport equations at 
$n\le10$ 
leads to a small underestimation of all low-lying states because of neglecting 
the de-excitation from the truncated ``tail'' with $n>10$. A possible influence 
of this cut has been estimated by comparing the above solution with a 
solution, cut at $n\le9$, where the de-excitation from states with $n=10$ 
are also truncated. Estimated in this way, the population of long-lived states 
at the Be exit is with a very conservatively chosen error band 
$(6.29^{+0.33}_{-0.13}) \cdot 10^{-2}$ \cite{DIRAC18}. 

An additional correction to this value may arise from the accuracy 
of atom-atom cross sections. 
The most accurate cross sections are obtained in the Glauber approach 
accounting for multi-photon exchange and the Hartree-Fock-Dirac form factors 
of the target atoms \cite{BASEl00,BASEl01,BASEl02}. 
As shown in \cite{AFAN03}, the accuracy of the used cross sections, 
compared to the most accurate ones, depends on the target material. 
In the case of the Be target, an estimation shows that the summed population of long-lived states for $n\leq8$ at the Be exit, calculated with the most accurate cross sections, is 2.8\% less than the current value. Thus, this value can be used as the scale factor accounting for uncertainties of the used atom-atom cross sections.  
Further, the accuracy of the most accurate cross sections  
is estimated to be about 1\% \cite{BASEl00,BASEl01,BASEl02}, 
which can be converted roughly to a 1\% error in the populations. 
Let us conservatively add this error linearly to the latter error band. 
Finally, the population of long-lived states at the Be exit is 
$(6.12^{+0.39}_{-0.19}) \cdot 10^{-2}$, compared to 
$(6.13^{+0.11}_{-0.09}) \cdot 10^{-2}$ in the above calculations. 
Thus, this crude accuracy estimation leads mainly to 
an increase of the error band for the calculated probabilities. 
Nevertheless, it is significantly smaller than the experimental error. 
Hence, its influence is applied only to the final results as 
described below. 

\subsection{Decay in gap and breakup in Pt for long-lived atoms}
\label{sec:gap}

In the 96~mm gap between the Be target and the Pt foil, the populations of 
the atomic states alter depending on their lifetime \cite{note1501}, 
calculated in QED. 
The decay length of atoms with fixed $n$
is minimal for $l=1$ and strongly increasing with $l$. 
Figure~\ref{fig:4} indicates that mainly states with $n=2-4$ are
decaying in the gap. The summed population of long-lived states at 
the entry of the Pt foil is $0.037\pm0.001$. 
Note that all cascade radiation transitions 
between atomic states as well as annihilation of $ns$ states are 
considered, using the technique of transport equations. 
The cascade accounting increases the summed population by about
10\% compared to a pure exponential decay. 

Table~\ref{tab:BrPt} shows the summed populations of long-lived states 
at the Be target exit and at the Pt foil entry as a function of $n$. 
Illustrative values of $np$ state breakup probabilities in the Pt foil 
are shown as well. The pionium breakup 
in Pt is treated in the same way as the breakup in the target. 
The breakup probability of $A^{L}_{2\pi}$ entering in Pt, 
averaged over populations, is 0.944. 

Finally, the expected total probability 
$P_\mathrm{br}^{\mathrm{tot}}(\mathrm{Pt})$, 
that pionium produced in Be will break up in Pt, 
is calculated via the difference between 
the total populations at the entry and exit of the Pt foil. 
For the average atom momentum (see above), this probability 
is found to be:
\begin{equation}
\label{eq:p_pt}
P_\mathrm{br}^{\mathrm{tot}}(\mathrm{Pt}) = 0.035\pm0.001 \; .
\end{equation}

\begin{table}[tb]
\caption{
Summed (over $l$ and $m$) populations of long-lived atomic states versus $n$. 
The populations at the Be target exit and at the Pt foil entry are given in \% of the total number of produced $A_{2\pi}$.
$P_\mathrm{br}^\mathrm{Pt}(np)$ is the breakup probability of 
the $A_{2\pi}$ $np$ states in the 2.1~\textmu{}m thick Pt foil. 
The values are calculated for the average atom momentum 
4.44~GeV/$c$ and the ground state lifetime $\tau=3.15\cdot10^{-15}$s. 
}
\label{tab:BrPt}
  \centering
\renewcommand{\arraystretch}{1.1}
{\small 
\begin{tabular}{@{}l@{\hspace{\tabcolsep}}lllllll@{}}
 \hline
$n$ & 2 & 3 & 4 & 5 & 6 & 7 & 8 \\
 \hline
Be & 2.38 & 1.56 & 0.85 & 0.49 & 0.30 & 0.20 & 0.13 \\
 \hline
Pt &0.48 & 1.10 & 0.76 & 0.47 & 0.30 & 0.19 & 0.13\\
 \hline
$P_\mathrm{br}^\mathrm{Pt}(np)$ & 0.763 & 0.933 & 0.978 & 0.991 & 0.996 & 
0.998 & 0.999\\
 \hline
\end{tabular} 
}
\end{table}


\section{Number of atoms produced in the Be target}
\label{sec:analysis_Be}

The analysis of $\pi^+\pi^-$ experimental data, produced in 
the Be target, is similar to the $\pi^+\pi^-$ analysis as 
presented in \cite{ADEV11}. The main difference is a shift in 
$Q_Y$ by 13.1 MeV/$c$, due to the magnetic field of 
the inserted permanent magnet. For the analysis in 
the current section, the found $Q_Y$ value of each 
experimental and MC event is lessened by the above 13.1 MeV/$c$. 

Events with transverse $Q_T < 4$~MeV/$c$ and 
longitudinal $|Q_L| < 15$~MeV/$c$ are selected to be analysed. 
The experimental $Q$-distribution $N(Q_i)$ and 
the distributions of its projections are fitted by 
simulated distributions of atomic $n^{MC}_A(Q_i)$, 
Coulomb $N^{MC}_{C}(Q_i)$ and non-Coulomb $N^{MC}_{nC}(Q_i)$ pairs. 
Using the difference of the particle production times, 
the admixture of accidental pairs is subtracted from 
the experimental distributions. The distributions of 
simulated events, $n^{MC}_A(Q_i)$), $N^{MC}_{C}(Q_i)$ and 
$N^{MC}_{nC}(Q_i)$, are normalized to 1. In the experimental 
distributions, the numbers of atomic ($n_A$), Coulomb ($N_C$) 
and non-Coulomb ($N_{nC}$) pairs are free fit parameters in 
the minimizing expression:

\[
\begin{split}
&\chi^2 =  \\
&\sum_i \frac{(N(Q_i) -
n_A \cdot n^{MC}_A(Q_i) -
N_C \cdot N^{MC}_{C}(Q_i) -
N_{nC} \cdot N^{MC}_{nC}(Q_i))^2}
{\sigma^{2}_{N(Q_i)}} .
\end{split}
\]
The sum of these parameters is equal to the number of analysed events. 
The fitting procedure takes into account the statistical errors of 
the experimental distributions.

\begin{figure}[tb]
\begin{center}
\includegraphics[width=0.9\columnwidth]{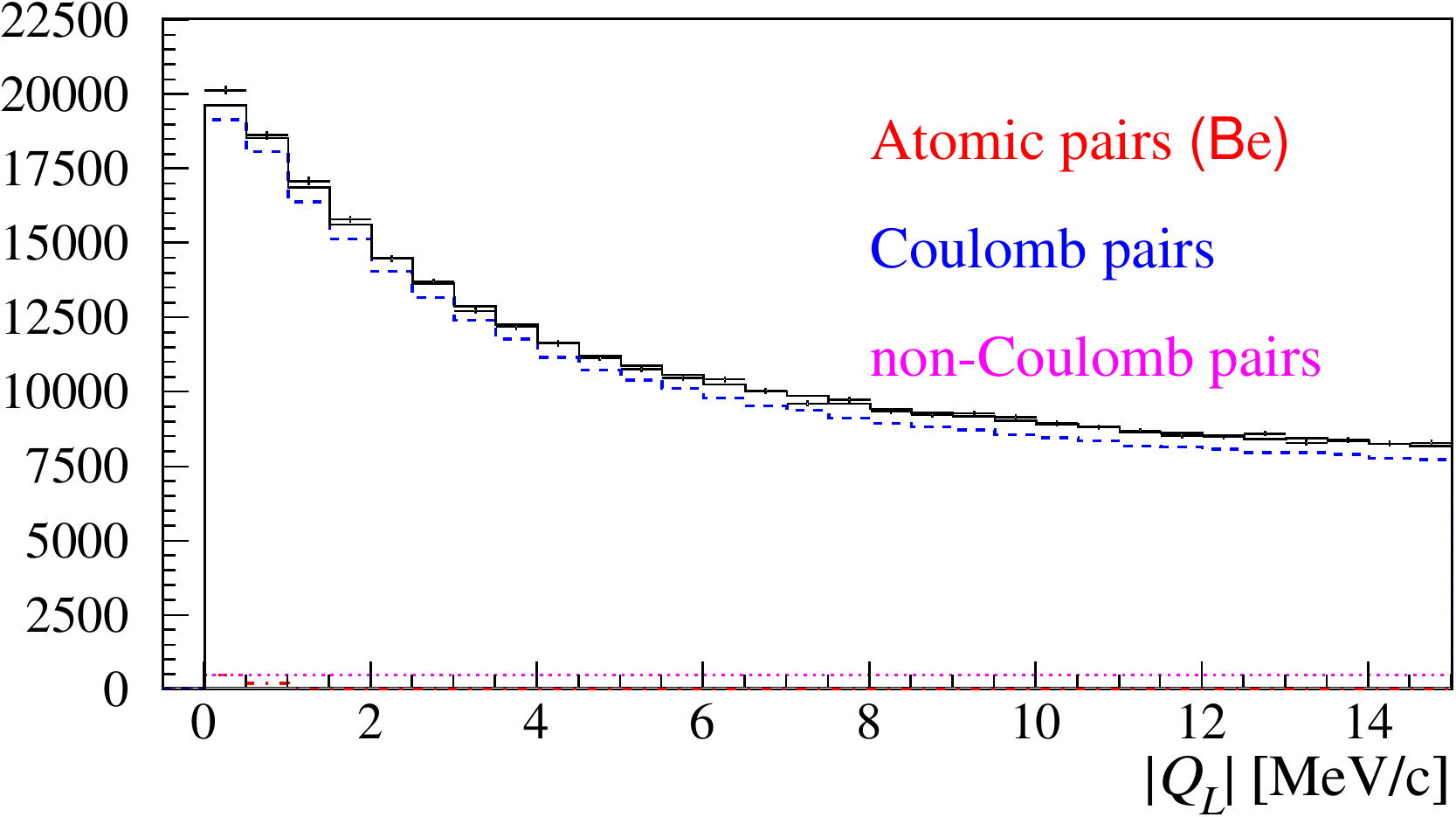}
\caption{Experimental $Q_L$ distribution of $\pi^+\pi^-$ pairs 
(+ with error bar) for the beryllium (Be) target fitted by 
the sum of simulated distributions of ``atomic'', ``Coulomb'' 
and ``non-Coulomb'' pairs. The summed distribution of free 
(``Coulomb'' and ``non-Coulomb'') pairs is shown as black line.} 
\label{fig:6}
\end{center}
\vspace{-5ex}
\end{figure}

Figure~\ref{fig:6} shows the experimental and 
simulated $Q_L$ distributions of $\pi^+\pi^-$ pairs for 
the data obtained from the Be target. One observes a very good agreement 
of the sum of simulated Coulomb and non-Coulomb pairs with experimental data 
in the region of $Q_L>1.5$~MeV/$c$, where no atomic pairs are expected. 
The main result of this fit is the number of Coulomb pairs: 
using (\ref{eq:number_A}), this number allows to obtain 
the full number of $\pi^+\pi^-$ atoms produced in the Be target. 

The Coulomb pairs from this distribution are used to reconstruct 
the lab momentum of produced atoms, shown in Figure~\ref{fig:3}. 
The average momentum is $\langle p_A \rangle=4.44~$GeV/$c$. This spectrum 
is used in the atom simulation. 

Table~\ref{tab:nA_Be} summarizes the number of Coulomb pairs $N_C$ from 
the 1-dimensional $Q$ and $|Q_L|$ analyses and from the 2-dimensi\-onal 
($|Q_L|$,$Q_T$) analysis. 
The efficiency of Coulomb pair recording is evaluated from 
the simulated data as ratio of the MC Coulomb pair number $N_C^{MC}$, 
passing the corresponding cuts --- in each of the above analyses --- to 
the full number of generated Coulomb pairs $N_C^{MC}(0)$: 
$\varepsilon_C=N_C^{MC}/N_C^{MC}(0)$. 
The full number of produced Coulomb pairs in the Be target is 
then given by $N_C/\varepsilon_C$.  
This number allows to calculate the number $N_A$ of atoms 
produced in the Be target: using the theoretical ratio 
$K=N_A/N_C$ (\ref{eq:number_A}), calculated for $q \le 2$~MeV/$c$, 
and the simulated probability 
$\varepsilon_K=N_C^{MC}(K)/N_C^{MC}(0)$, that the Coulomb pairs 
have momenta $q \le 2$~MeV/$c$, results in 
$N_A=K\cdot\varepsilon_K\cdot N_C/\varepsilon_C$.  
To recover the losses due to the coplanarity cut in 
the trigger system (Sec.~\ref{sec:setup}), 
the number $N_A$ is additionally corrected (MC). 
The corrected values $N^\mathrm{tot}_A$ are shown in the last column of 
Table~\ref{tab:nA_Be}.

\begin{table}[htb]
\caption{Number $N_C$ of Coulomb pairs from different analyses and  
fit quality as $\chi^{2}/n$ ($n$ is degree of freedom).  
$N_A$ is the corresponding number of atoms produced in the Be target 
and $N^\mathrm{tot}_A$ the total number of atoms after correction for 
the coplanarity cut. The errors are statistical ones.}
\label{tab:nA_Be}
\renewcommand{\arraystretch}{1.2}
{\small 
\begin{tabular}{@{}l@{\hspace{\tabcolsep plus \tabcolsep}}llll@{}}
\hline
Analysis & $N_C$ & $\chi^2/n$ & $N_A$ & $N^\mathrm{tot}_A$   \\ \hline
$Q$ & $321340\pm2660$ & 33/27& $12539 \pm 97$  & $17000 \pm 130$  \\ 
$|Q_L|$ & $315830\pm3250$ & 30/27 & $12320 \pm 120$ & $16700 \pm 160$ \\ 
$|Q_L|,Q_T$ & $319890\pm2610$ & 152/117 & $12483 \pm 97$  & $16960 \pm 130$ \\
\hline
\end{tabular}
}
\end{table}


\section{Number of long-lived atoms broken up in Pt} 
\label{sec:analysis_Pt}

Atomic pairs from the $A_{2\pi}^L$ atom breakup in the Pt foil are 
observed \cite{ADEV15} above background $\pi^+\pi^-$ pairs 
produced mainly in the Be target. For the analysis, the same approach 
is applied as in Section~\ref{sec:analysis_Be}, but this time 
$Q_Y$ is shifted by 2.3 MeV/$c$ for all pairs instead of 13.1 MeV/$c$
(Figure~\ref{fig:1}). 

\begin{figure}[tb]
\includegraphics[width=0.9\columnwidth]{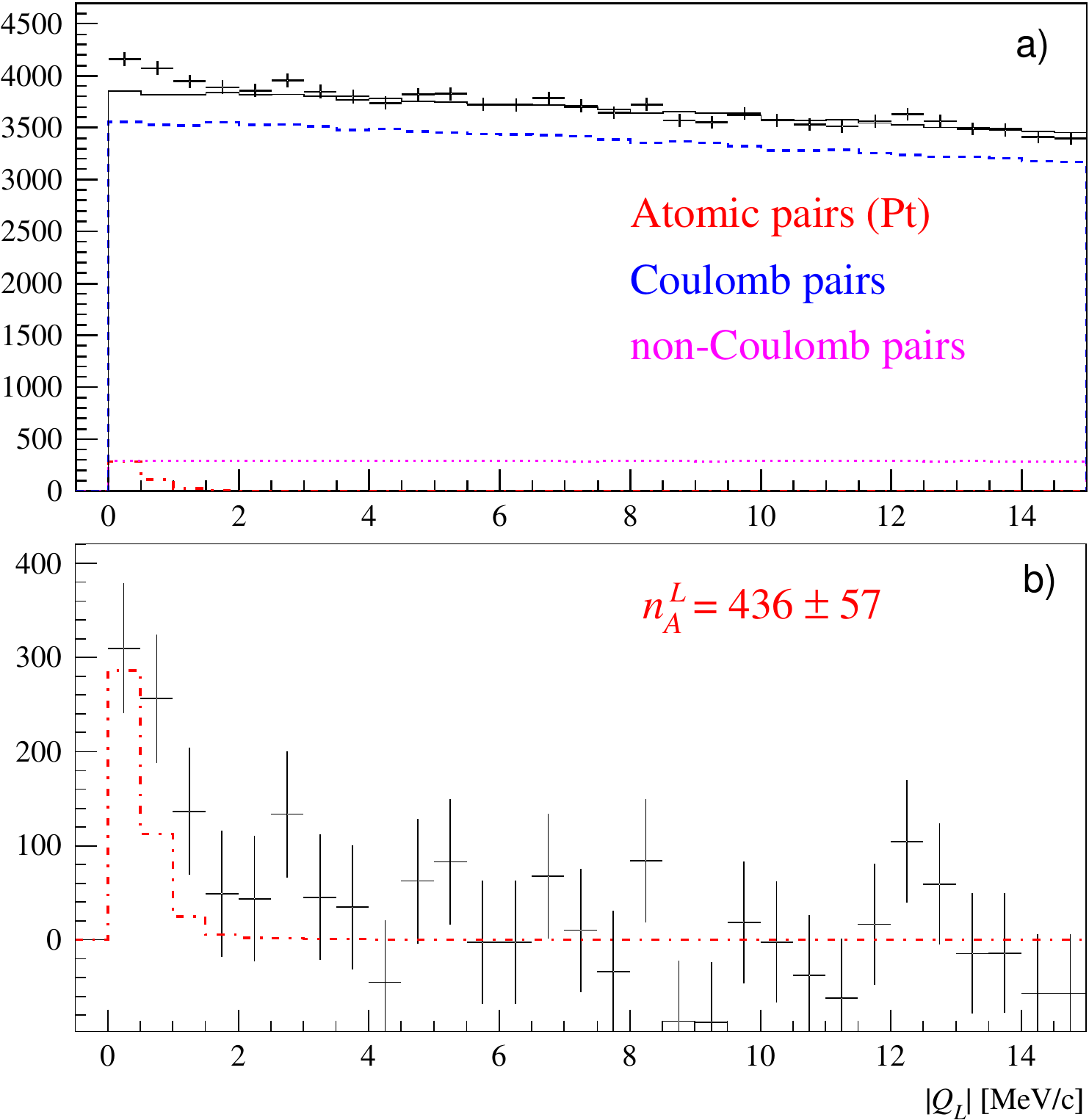}
\caption{$|Q_L|$ distribution of $\pi^+\pi^-$ pairs for 
$Q_T<2.0$~MeV/$c$. Plot a) presents the experimental distribution 
(+ with statistical error bar) and the simulated background (solid line).
Plot b) shows the experimental distribution after 
background subtraction (+ with statistical error bar) and 
the simulated distribution of atomic pairs from $A_{2\pi}^L$ 
broken up in Pt (dotted-dashed line). The fit procedure has been 
applied to the 2-dimensional ($|Q_L|,Q_T$) distribution.}
\label{fig:7}
\end{figure}

In the 2-dimensional ($|Q_L|,Q_T$) analysis, experimental 
data are analysed by means of corresponding simulated distributions. 
For $|Q_L|<15$~MeV/$c$ and $Q_T<2$~MeV/$c$, 
the $|Q_L|$ projection of the experimental 2-dimensional 
distribution as well as of the three types of simulated 
$\pi^+\pi^-$ pairs is shown in Figure~\ref{fig:7}a. 
One observes an excess of events --- above the sum of Coulomb and 
non-Coulomb pairs --- in the low $Q_L$ region, where atomic pairs 
are expected. After background subtraction, there is 
a statistically significant signal of $n_A^L=436 \pm 57$ 
as presented in Figure~\ref{fig:7}b. The signal shape is 
compared with the simulated distribution of atomic pairs 
resulting from the $A_{2\pi}^L$ breakup in the Pt foil. 
The description is acceptable in view of large 
statistical uncertainties induced by subtracting two large numbers 
in the bins of the signal distribution (discussion below). 

The $\Lambda$ width correction accuracy (Section~\ref{sec:setup}) 
leads to a systematic error in $n_A^L$ of $\sigma^{syst}_{\Lambda}=4.4$. 
The accuracy of the measured Pt foil thickness causes
a systematic error of $\sigma^\mathrm{syst}_{Pt}=22$ in 
the 2-dimensional analysis. Generation of $\pi^+\pi^-$ pairs by beam 
halo protons is negligible \cite{ADEV15}. 
                                              
An additional 1-dimensional $|Q_L|$ analysis is performed 
to check the influence of the simulated atomic pair shape. 
The experimental distribution is analysed with  
3 free parameters (fractions of atomic, Coulomb and 
non-Coulomb pairs) in the range $0<|Q_L|<15$~MeV/$c$, and 
with 2 parameters (fractions of Coulomb and non-Coulomb pairs) in 
the interval $2<|Q_L|<15$~MeV/$c$, where no atomic pairs are expected. 
In the region $|Q_L|<2$, $Q_T<4$~MeV/$c$, 
the atomic pair numbers obtained with the 3-parameter fit are 
$n^L_A = 435 \pm 103$ and obtained with the 2-parameter fit 
$n^L_A = 579 \pm 164$. The difference in 
$n^L_A$ is mainly due to different signal shapes in the experiment 
and simulation and is taken as 
an estimate for the systematic error (only with positive sign) 
$\sigma^{syst}_{shape}=144$. The systematic error of 
the number of atoms produced in Be ($N^{\mathrm{tot}}_A$) is calculated 
as a maximum difference between the value of the analysis with the 
current variables and other analyses, $\sigma^{syst}_{A}=260$. 

In summary, the analysis of the 2-dimensional ($|Q_L|,Q_T$) 
distributions yields the following value for the total probability 
that a $\pi^+\pi^-$ atom produced in Be breaks up in the Pt foil:

\begin{equation}\label{eq:pBr}
\begin{split}
P_\mathrm{br}^{\mathrm{tot}}(\mathrm{Pt}) & = 
\frac{n_A^L}{N_A^\mathrm{tot}}=\left. 0.0257 \pm 0.0034\right|_\mathrm{stat} 
\left.^{+0.0086}_{-0.0014}\right|_\mathrm{syst} \\
& = \left. 0.0257^{+0.0092}_{-0.0036}\right|_\mathrm{tot} .
\end{split}
\end{equation}

\section{Measuring a long-lived $\pi^+\pi^-$ atom lifetime}
\label{sec:analysis}

A long-lived atom lifetime is evaluated by means of two analyses. 
In the first simplified analysis, all long-lived atoms are considered as 
objects, which have one common lifetime, the average momentum 
$\langle P_A \rangle$ and the breakup probability in the Pt foil equal to 1.
The long-lived atom lifetime estimated in this way is
$$
\label{eq:tmin}
\tau=
\left(\left.2.32\pm0.35|_\mathrm{stat}
{}^{+0.90}_{-0.15}\right|_\mathrm{syst}\right) 
\cdot10^{-11}\:\mathrm{s} =
\left(\left.2.32^{+0.96}_{-0.38}\right|_\mathrm{tot}\right) 
\cdot10^{-11}\:\mathrm{s}.
$$
This value of the long-lived atom lifetime is more than three orders of magnitude 
larger than our measured value of the atom ground state lifetime 
$\tau_{1s} = \left( \left. 3.15^{+0.28}_{-0.26} \right|_\mathrm{tot} \right) 
\cdot 10^{-15}$~s \cite{ADEV11}. 
Accounting for the additional calculation errors, 
discussed in Section~\ref{sec:ll-pop}, leads to almost the same value: 
$\tau=(\left.2.33^{+0.98}_{-0.39}\right|_\mathrm{tot}) \cdot10^{-11}$~s. 

In the second more sophisticated analysis, the populations of 
the pionium states are described in terms of transport equations 
for the whole path from the production point in the Be target to 
the exit of the Pt foil. 
For the gap between Be and Pt, 
radiation transition rates and annihilation of short-lived states 
are included in the transport equations. The variation of  
the different $A_{2\pi}^L$ lifetimes $\tau_i$ from their QED values 
$\tau_i^{QED}$ \cite{note1501} is done via one common factor 
$\alpha=\tau_i/\tau_i^{QED}$. All the probabilities involved in 
these calculations are averaged over 
the spectrum of the observed atoms (Figure~\ref{fig:3}). 

\begin{figure}[tb]
\begin{center}
\includegraphics[width=0.9\columnwidth]{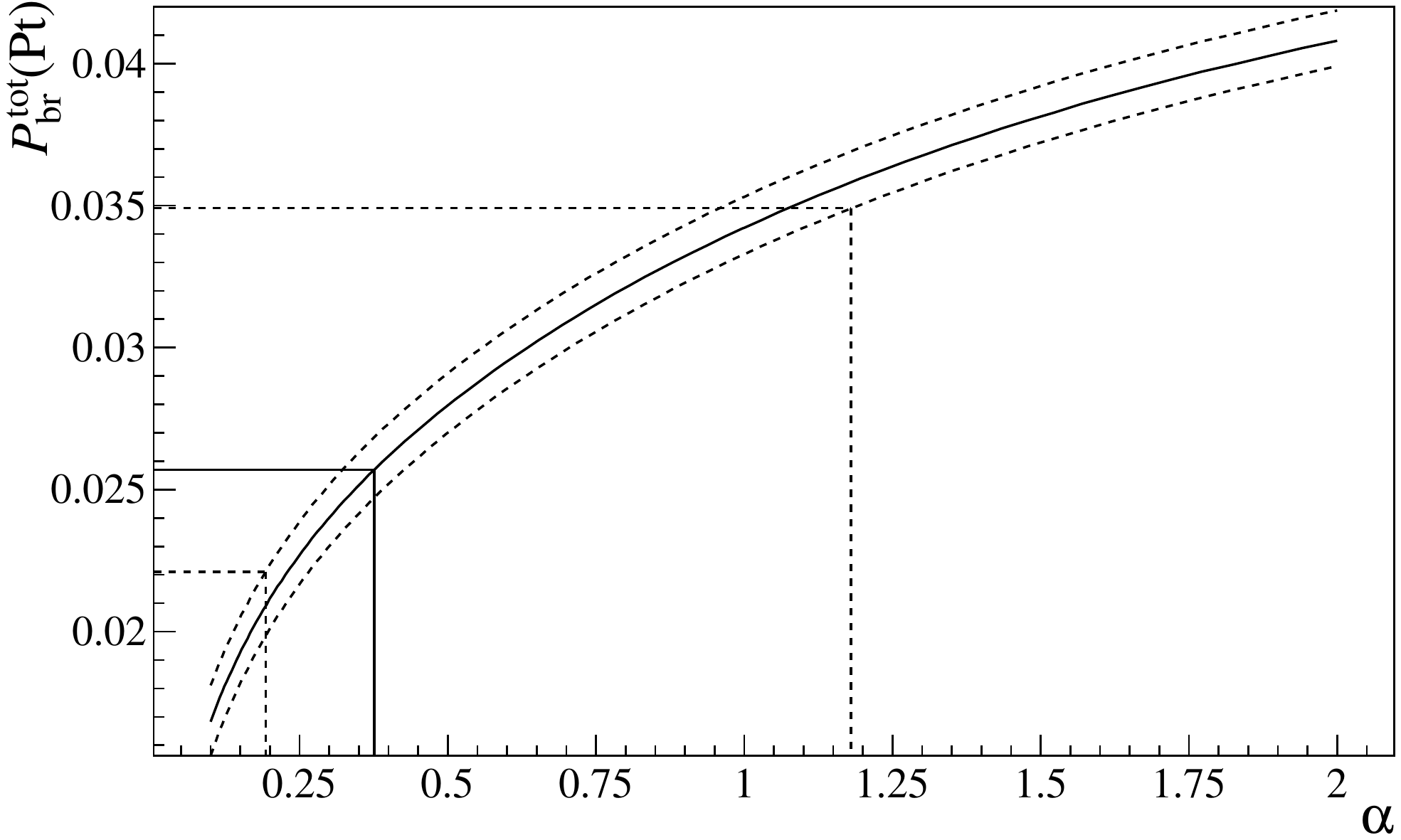}
\end{center}
\vspace{-3ex}
\caption{
Probability $P_\mathrm{br}^{\mathrm{tot}}(\mathrm{Pt})$ calculated as 
a function of $\alpha$ (see text). 
The horizontal lines correspond to the measured value 
$P_\mathrm{br}^{\mathrm{tot}}(\mathrm{Pt}) = 
\left. 0.0257^{+0.0097}_{-0.0036}\right|_\mathrm{tot}$ 
(\ref{eq:pBr}) together with the total errors. 
The value $\alpha=1$, which corresponds to pure QED calculations, 
is within the error band of the measurement.}	
\label{fig:8}
\vspace{-2ex}
\end{figure}

The solid line curve in Figure~\ref{fig:8} presents the breakup probability  
$P_\mathrm{br}^{\mathrm{tot}}(\mathrm{Pt})$ (see (\ref{eq:p_pt})) 
as a function of the factor $\alpha$. The error band (dashed curves) 
accounts for the different extrapolations of the state population for 
$n\ge10$ (Section~\ref{sec:long-lived}). The horizontal lines are 
the measured value $P_\mathrm{br}^{\mathrm{tot}}(\mathrm{Pt})$  
together with the total errors (\ref{eq:pBr}). This value corresponds 
to $\alpha= 0.376^{+0.804}_{-0.183}$. As $\alpha=1$ is included in 
the error band, one concludes that the measured lifetime does not 
contradict the QED calculations. The lifetime of the $2p$~state, 
which is the shortest-lived of all long-lived states, is found to be 
$\tau_{2p}=\left(\left.0.44^{+0.94}_{-0.21}\right|_\mathrm{tot}\right)
\cdot10^{-11}$~s. This value is in agreement with the calculation in QED, 
$\tau_{2p}=1.17 \cdot 10^{-11}$~s. 

Accounting for the additional calculation errors, discussed in 
Section~\ref{sec:ll-pop}, leads to a wider error band for 
$\alpha= 0.383^{+0.926}_{-0.254}$ and the $2p$ lifetime $\tau_{2p}$: 
\begin{equation}
\label{eq:t2p}
\tau_{2p}=\left(\left.0.45^{+1.08}_{-0.30}\right|_\mathrm{tot}\right)
\cdot10^{-11}\mathrm{s} \, .
\end{equation} 

The magnetic field between the Be target and the Pt foil  
transforms into an electric field in the atom c.m. system. 
This field mixes the wave functions of long-lived $np$ states 
and short-lived $ns$ states resulting in shortening 
the observed lifetime of $np$ states. The influence of 
the magnetic field on the lifetime for $2p$ and $2s$ states 
has been evaluated in \cite{NEME01}. The magnetic field in 
the gap shortens the $2p$ lifetime by 1.002 and even more 
for higher $n$. To estimate a maximum reduction, all $np$ states 
with $n>3$, around 8.6\% of all long-lived states at the Be exit, 
have been considered as decayed in the gap. 
In this extreme case, the value of the $2p$ lifetime is 
$\tau_{2p}=(\left.0.60^{+1.34}_{-0.30}\right|_\mathrm{tot}) \cdot10^{-11}$~s, 
not contradicting the result in (\ref{eq:t2p}), evaluated without 
taking into account the influence of the magnetic field.

\section{Conclusion}

The evaluated lifetime 
$\tau_{2p}=(\left.0.45^{+1.08}_{-0.30}\right|_\mathrm{tot}) \cdot10^{-11}$~s 
is three orders of magnitude larger than our previously measured value of 
the $A_{2\pi}$ ground state lifetime 
$\tau=(\left.3.15^{+0.28}_{-0.26}\right|_\mathrm{tot})\cdot 10^{-15}$s  
\cite{ADEV11}. This outcome opens a possibility 
to measure the \textit{Lamb shift} of this atom and, 
herewith, a new $\pi\pi$ scattering length combination $2a_0+a_2$. 
These scattering lengths have been calculated in the framework of 
Chiral Perturbation Theory and are nowadays investigated in Lattice QCD.

\section*{Acknowledgements}

We are grateful to O.V.~Teryaev for the useful discussions, A.~Vorozhtsov, 
D.~Tommasini and their colleagues from TE-MSC/CERN for the Sm-Co magnet design 
and construction, R.~Steerenberg and the CERN-PS crew for 
the delivery of a high quality proton beam and 
the permanent effort to improve the beam characteristics. 
The project DIRAC has been supported by 
CERN, the JINR administration, the Ministry of Education and 
Youth of the Czech Republic by project LG130131, 
the Istituto Nazionale di Fisica Nucleare and the University of Messina (Italy),  
the Grant-in-Aid for Scientific Research from 
the Japan Society for the Promotion of Science, 
the Ministry of Research and Innovation (Romania), 
the Ministry of Education and Science of the Russian Federation and 
Russian Foundation for Basic Research, 
the Direcci\'{o}n Xeral de Investigaci\'{o}n, Desenvolvemento e Innovaci\'{o}n, 
Xunta de Galicia (Spain) and the Swiss National Science Foundation.

\end{document}